\documentclass[aps,twocolumn,showpacs]{revtex4}
\usepackage{epsf}

\begin{document}
% IISc-CHEP-6/05

\title{Quantum Random Walks without Coin Toss%
\footnote{Invited lecture at the Workshop on Quantum Information,
          Computation and Communication (QICC-2005), IIT Kharagpur,
          India, February 2005, {\tt quant-ph/0506221}.}
}
\author{Apoorva Patel}
\affiliation{\rm (Collaborators: K.S. Raghunathan and Pranaw Rungta)\\
             Centre for High Energy Physics,\\
             Indian Institute of Science, Bangalore-560012, India\\
             E-mail: adpatel@cts.iisc.ernet.in}

\begin{abstract}
\noindent
We construct a quantum random walk algorithm, based on the Dirac operator
instead of the Laplacian. The algorithm explores multiple evolutionary
branches by superposition of states, and does not require the coin toss
instruction of classical randomised algorithms. We use this algorithm to
search for a marked vertex on a hypercubic lattice in arbitrary dimensions.
Our numerical and analytical results match the scaling behaviour of earlier
algorithms that use a coin toss instruction.
\end{abstract}
\maketitle

\section{Introduction}

Random walks are a fundamental ingredient of non-deterministic algorithms
\cite{motwani},
and are used to tackle a wide variety of problems---from graph structures
to Monte Carlo samplings. Such algorithms have many evolutionary branches,
which are explored probabilistically, to estimate the correct result.
A classical computer can explore only one branch at a time, so the algorithm
is executed several times, and the estimate of the final result is extracted
from the ensemble of individual executions by methods of probability theory.
Such algorithms are typically represented using graphs, with vertices denoting
the states and the edges denoting the evolutionary routes. A particular
evolution corresponds to a specific walk on the graph, and the final result
is obtained by combining the results for many different walks. To ensure that
different evolutionary branches are explored in different executions, one
needs non-deterministic instructions, and they are provided in the form of
random numbers. A coin toss is the simplest example of a random number
generator, and it is included in the instruction set for a probabilistic
Turing machine.

A quantum computer can explore multiple branches of a non-deterministic
algorithm in a single attempt, by using a clever superposition of states.
The probabilistic result can then be arrived at by interference of amplitudes
corresponding to different branches. Thus as long as the means to construct
a variety of superposed states exist, there is no a priori reason to include
a coin toss as an instruction for a (probabilistic) quantum Turing machine.

In what follows, we construct a quantum random walk on a hypercubic lattice
in arbitrary dimensions without using a coin toss instruction, analyse its
properties, and use it to find a marked vertex on the lattice. More details
are available in Refs.\cite{qwalk1,qwalk2}.

\section{Diffusion}

A random walk is a diffusion process, commonly described using the Laplacian
operator in the continuum. To construct a discrete quantum random walk,
we must discretise the diffusion process using evolution operators that are
both unitary and ultra-local (an ultra-local operator vanishes outside a
finite range).

On a periodic lattice, the spatial modes are characterised by discrete wave
vectors $\vec{k}$. Quantum diffusion then depends on the energy of these
modes according to $U(\vec{k},t) = \exp(-iE(\vec{k})t)$. The lowest energy
mode, $\vec{k}=0$, corresponding to a uniform distribution, is an eigenstate
of the diffusion operator and does not propagate. The slowest propagating
modes are the ones with smallest nonzero $|\vec{k}|$. The classical Laplacian
operator gives $E(\vec{k}) \propto|\vec{k}|^2$ \cite{massterm},
which translates to the characteristic Brownian motion signature, spread
$\langle n \rangle_{\rm rms} \propto \sqrt{t}$. There is an alternative in
quantum theory---instead of the non-relativistic Sch\"odinger equation based
on the Laplacian operator $\nabla^2$, one can use the relativistic Dirac
equation based on the operator $\nabla\!\!\!\!/$. The Dirac operator gives
$E(\vec{k})\propto|\vec{k}|$, with the associated signature, spread
$\langle n \rangle_{\rm rms} \propto t$. Clearly the Dirac operator, with
its faster diffusion of the slowest modes compared to the Laplacian, is the
operator of choice for constructing faster diffusion based quantum algorithms.

An automatic consequence of the Dirac operator is the appearance
of an additional internal degree of freedom corresponding to spin,
whereby the quantum state is described by a multi-component spinor.
These spinor components were identified with the states of a coin in
Refs.\cite{gridsrch1,gridsrch2},
with the coin evolution rule guiding the quantum diffusion process.
While this is the correct procedure in the continuum theory, another option
is available for a lattice theory, i.e. staggered fermions \cite{staggered}.
In this approach, the spinor degrees of freedom are spread out over an
elementary hypercube, location dependent signs appear in the evolution
operator, and translational invariance exists in steps of 2 instead of 1.
We follow this approach to construct, a quantum diffusion process on a
hypercubic lattice, without a coin toss instruction,

The free particle Dirac Hamiltonian in $d$-space dimensions is
\begin{equation}
H_{\rm free} = -i\vec{\alpha}\cdot\vec{\nabla} + \beta m ~.
\end{equation}
On a hypercubic lattice, the simplest discretisation of the derivative
operator is
\begin{equation}
\nabla_k f(\vec{x}) = {1 \over 2}[ f(\vec{x}+\hat{k}) - f(\vec{x}-\hat{k}) ] ~.
\end{equation}
Then the anticommuting matrices $\vec{\alpha},\beta$ can be
spin-diagonalised to the location dependent signs
\begin{equation}
\alpha_k = \prod_{j=1}^{k-1} (-1)^{x_j} ~,~~
\beta = \prod_{j=1}^{d} (-1)^{x_j} ~.
\end{equation}

\begin{figure}[b]
\epsfxsize=8cm
\centerline{\epsfbox{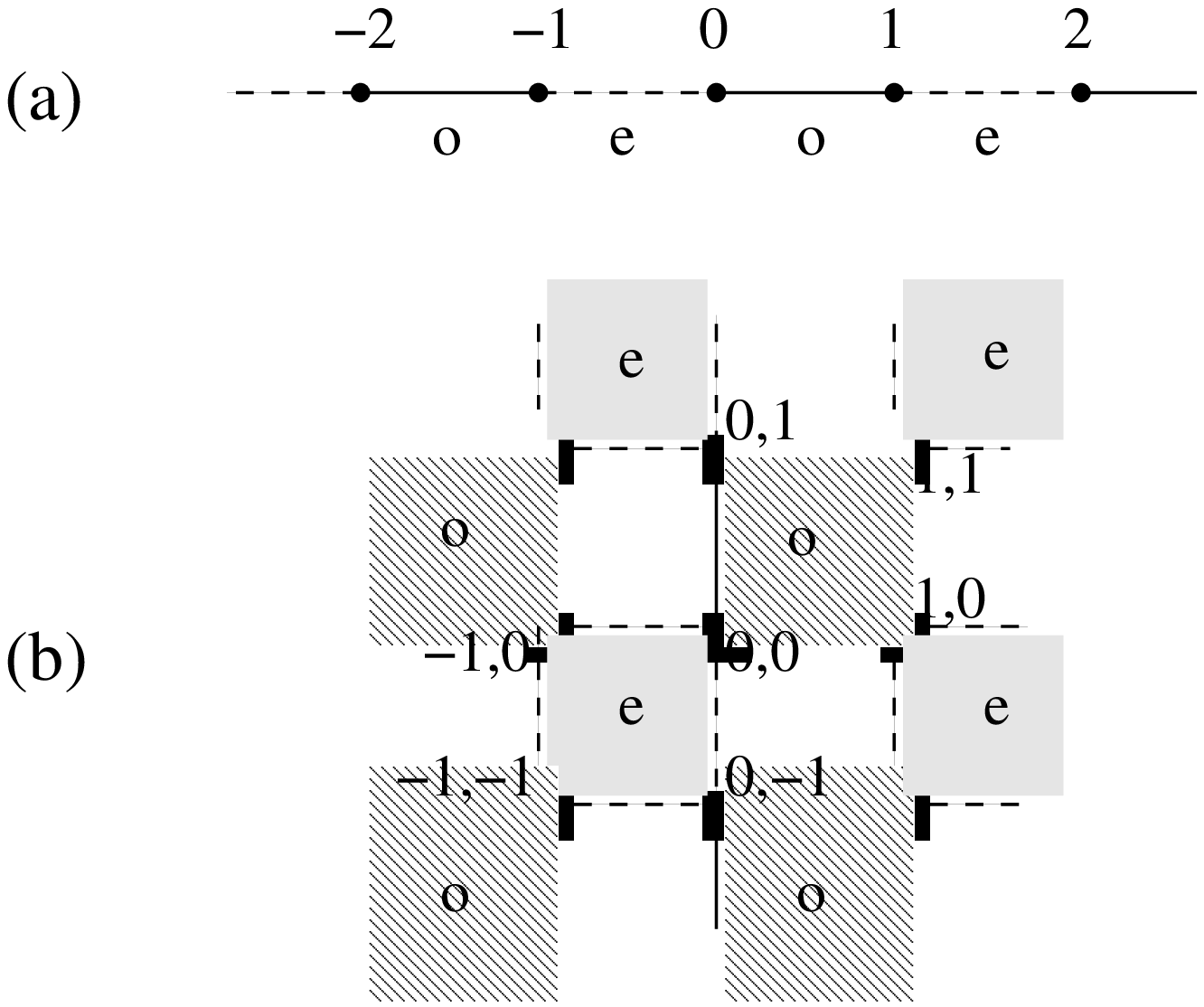}}
\caption{Partitioning of lattice in to odd and even parts for ultra-local
discrete quantum walk: (a) for $d=1$, (b) for $d=2$.}
\end{figure}

Even when the Hamiltonian $H$ is ultra-local (i.e. has a finite range),
the evolution operator $U=\exp(-iHt)$ is not. To make the evolution
operator ultra-local, we break up $H$ in to block-diagonal Hermitian parts,
and then exponentiate each part separately.
Partitioning of $H$ in to two parts (which we label ``odd'' and ``even'')
is sufficient for this purpose \cite{spinorsize}.
This partition is illustrated in Fig.1 for $d=1$ and $d=2$. Each part
contains all the vertices but only half of the links attached to each vertex.
Consequently, each link appears in only one of the two parts, and can be
associated with a term in $H$ providing propagation along it, i.e.
\begin{equation}
H_{\rm free} = H_o + H_e ~.
\end{equation}
The Hamiltonian is thus divided in to a set of non-overlapping blocks
that can be exponentiated exactly. Each block is an elementary hypercube
on the lattice, and the block matrices are of size $2^d\times2^d$ in $d$
dimensions. The ultra-local quantum random walk on the lattice then
evolves the amplitude distribution according to
\begin{equation}
\psi(\vec{x};t) = W^t \psi(\vec{x};0) ~,~~
\end{equation}
\begin{equation}
W = U_e U_o = e^{-iH_e\tau} e^{-iH_o\tau} ~.
\end{equation}
Each block of the unitary matrices $U_{o(e)}$ mixes the amplitudes of
vertices belonging to a single elementary hypercube, and the amplitude
distribution spreads because the two alternating matrices do not commute.
The random walk operator $W$ is translationally invariant in steps of 2,
along each coordinate direction.

\section{Quantum Random Walk on a Line}

\subsection{Construction}

To explicitly illustrate the above described procedure, let us consider
the random walk on a line, with the allowed positions labeled by integers.

The simplest translation invariant ultra-local discretisation of the
Laplacian operator is
\begin{equation}
H |n\rangle ~\propto~ \big[ -|n-1\rangle + 2|n\rangle - |n+1\rangle \big] ~.
\end{equation}
One may search for ultra-local translationally invariant unitary evolution
operators using the ansatz
\begin{equation}
U |n\rangle ~=~ a|n-1\rangle + b|n\rangle + c|n+1\rangle ~,
\end{equation}
but then the orthogonality constraints between different rows of the unitary
matrix make two of $\{a,b,c\}$ vanish, and one obtains a directed walk
instead of a random walk. This problem can be bypassed, and an ultra-local
unitary random walk can be constructed, by enlarging the Hilbert space with
a quantum coin, e.g.
\begin{equation}
U = \sum_n \Big[
    |\!\uparrow\rangle\langle\uparrow\!|     \otimes |n+1 \rangle\langle n|
  + |\!\downarrow\rangle\langle\downarrow\!| \otimes |n-1 \rangle\langle n|
           \Big] .
\label{evolcoin}
\end{equation}
This route \cite{nayak}
brings its own set of caveats, due to quantum entanglement between the coin
and the position degrees of freedom.

We follow an alternate route familiar to lattice field theorists
\cite{staggered}.
It has also been used to simulate quantum scattering with ultra-local
operators \cite{richardson},
and to construct quantum cellular automata \cite{meyer}.
The starting point is the decomposition of the Laplacian operator in to its
even and odd parts, $H = H_e + H_o$,
\begin{equation}
H \propto \left(\matrix{
    \cdots&\cdots  &\cdots  &\cdots  &\cdots  &\cdots  &\cdots  &\cdots \cr
    \cdots&\hfill-1&\hfill 2&\hfill-1&\hfill 0&\hfill 0&\hfill 0&\cdots \cr
    \cdots&\hfill 0&\hfill-1&\hfill 2&\hfill-1&\hfill 0&\hfill 0&\cdots \cr
    \cdots&\hfill 0&\hfill 0&\hfill-1&\hfill 2&\hfill-1&\hfill 0&\cdots \cr
    \cdots&\hfill 0&\hfill 0&\hfill 0&\hfill-1&\hfill 2&\hfill-1&\cdots \cr
    \cdots&\cdots  &\cdots  &\cdots  &\cdots  &\cdots  &\cdots  &\cdots \cr
    }\right) ,
\end{equation}
\begin{equation}
\!H_e \propto \left(\matrix{
    \cdots&\cdots  &\cdots  &\cdots  &\cdots  &\cdots  &\cdots  &\cdots \cr
    \cdots&\hfill-1&\hfill 1&\hfill 0&\hfill 0&\hfill 0&\hfill 0&\cdots \cr
    \cdots&\hfill 0&\hfill 0&\hfill 1&\hfill-1&\hfill 0&\hfill 0&\cdots \cr
    \cdots&\hfill 0&\hfill 0&\hfill-1&\hfill 1&\hfill 0&\hfill 0&\cdots \cr
    \cdots&\hfill 0&\hfill 0&\hfill 0&\hfill 0&\hfill 1&\hfill-1&\cdots \cr
    \cdots&\cdots  &\cdots  &\cdots  &\cdots  &\cdots  &\cdots  &\cdots \cr
    }\right) ,
\end{equation}
\begin{equation}
\!H_o \propto \left(\matrix{
    \cdots&\cdots  &\cdots  &\cdots  &\cdots  &\cdots  &\cdots  &\cdots \cr
    \cdots&\hfill 0&\hfill 1&\hfill-1&\hfill 0&\hfill 0&\hfill 0&\cdots \cr
    \cdots&\hfill 0&\hfill-1&\hfill 1&\hfill 0&\hfill 0&\hfill 0&\cdots \cr
    \cdots&\hfill 0&\hfill 0&\hfill 0&\hfill 1&\hfill-1&\hfill 0&\cdots \cr
    \cdots&\hfill 0&\hfill 0&\hfill 0&\hfill-1&\hfill 1&\hfill 0&\cdots \cr
    \cdots&\cdots  &\cdots  &\cdots  &\cdots  &\cdots  &\cdots  &\cdots \cr
    }\right) .
\end{equation}
While $H$ has the structure of a second derivative, its two parts,
$H_e$ and $H_o$, have the structure of a first derivative. The above
decomposition is indeed reminiscent of the ``square-root'' one takes
to go from the Laplacian to the Dirac operator.

The two parts, $H_e$ and $H_o$, are individually Hermitian. They are
block-diagonal with a constant $2\times2$ matrix, and so they can be
exponentiated while maintaining ultra-locality. The total evolution
operator can therefore be easily truncated, without giving up either
unitarity or ultra-locality,
\begin{eqnarray}
U(\Delta t) = e^{i(H_e+H_o)\Delta t} &=&
e^{iH_e\Delta t} e^{iH_o\Delta t} + O((\Delta t)^2) \\
&=& U_e(\Delta t) U_o(\Delta t) + O((\Delta t)^2) ~. \nonumber
\end{eqnarray}
The quantum random walk can now be generated using $U_e U_o$ as the
evolution operator for the amplitude distribution $\psi(n,t)$,
\begin{equation}
\psi(n,t) = [U_e U_o]^t \psi(n,0) ~,
\label{walkt}
\end{equation}
The fact that $U_e$ and $U_o$ do not commute with each other is enough
for the quantum random walk to explore all possible states. The price
paid for the above manipulation is that the evolution operator is
translationally invariant along the line in steps of 2, instead of 1.

The $2\times2$ matrix appearing in $H_e$ and $H_o$ is proportional to
$(1-\sigma_1)$, and so its exponential will be of the form $(c1+is\sigma_1)$,
$|c|^2+|s|^2=1$. A random walk should have at least two non-zero entries
in each row of the evolution operator. Even though our random walk treats
even and odd sites differently by construction, we can obtain an unbiased
random walk, by choosing the $2\times2$ blocks of $U_e$ and $U_o$ as
$\frac{1}{\sqrt2}{1\,i \choose i\,1}$. Furthermore, it is computationally
more convenient to choose a basis where the unitary operators are all real.
Performing a global phase transformation, $|n\rangle \rightarrow i^n|n\rangle$
\cite{phaseshift},
the $2\times2$ blocks of $U_e$ and $U_o$ become
$\frac{1}{\sqrt2}(1 \pm i\sigma_2)$. The discrete quantum random walk then
evolves the amplitude distribution according to
\begin{eqnarray}
\label{evolnocoin_o}
U_o |n\rangle &=& {1 \over \sqrt{2}}\Big[ |n\rangle - (-1)^n |n+(-1)^n\rangle \Big] , \\
\label{evolnocoin_e}
U_e |n\rangle &=& {1 \over \sqrt{2}}\Big[ |n\rangle + (-1)^n |n-(-1)^n\rangle \Big] ,
\end{eqnarray}
\begin{equation}
U_e U_o |n\rangle = {1 \over 2}\Big[
    |n-1\rangle + |n\rangle - |n+1\rangle + |n+2(-1)^n\rangle \Big] .
\label{walkstep}
\end{equation}

It is instructive to realise that, with the above choice, the unbiased quantum
random walk represents the path integral for a relativistic particle with
$|p|=m$. Its speed (in units of speed of light) is then $|v|=1/\sqrt{2}$.
The directed walk, with the $2\times2$ block matrix $U\propto\sigma_2$,
corresponds to $|v|=1$, and the stationary limit $U=1$ corresponds to $v=0$.

\subsection{Analysis}

It is straightforward to analyse the properties of the walk in
Eq.(\ref{walkstep}) using the Fourier transform:
\begin{equation}
\widetilde\psi(k,t) = \sum_n e^{ikn} \psi(n,t) ~,~
\end{equation}
\begin{equation}
\psi(n,t) = \int_{-\pi}^{\pi} {dk \over 2\pi}~e^{-ikn} \widetilde\psi(k,t) ~.
\end{equation}
The evolution of the amplitude distribution in Fourier space is easily
obtained by splitting it in to its even and odd parts:
\begin{equation}
\psi \equiv \left(\matrix{ \psi_e \cr \psi_o \cr}\right) ~,~~
\psi(k,t) = [M(k)]^t \psi(k,0) ~,~
\end{equation}
\begin{equation}
M(k) = \left(\matrix{ e^{ik}\cos k & -i\sin k      \cr
                      -i\sin k     & e^{-ik}\cos k \cr} \right) ~.
\end{equation}
The unitary matrix $M$ has the eigenvalues,
$\lambda_\pm \equiv e^{\pm i\omega_k}$
(this $\pm$ sign label continues in all the results that follow),
\begin{equation}
\lambda_\pm = \cos^2 k \pm i\sin k \sqrt{1+\cos^2 k} ~,~~
\omega_k = \cos^{-1}(\cos^2 k) ~,
\end{equation}
with the (unnormalised) eigenvectors,
\begin{eqnarray}
e_\pm &\propto& \left(\matrix{ -\cos k \mp \sqrt{1+\cos^2 k} \cr 1 \cr} \right)
~, \nonumber\\
      &\propto& \left(\matrix{ 1 \cr \cos k \mp \sqrt{1+\cos^2 k} \cr} \right)
~.
\end{eqnarray}
The evolution of amplitude distribution then follows
\begin{equation}
\widetilde\psi(k,t) = e^{ iw_k t} \widetilde\psi_+(k,0)
                    + e^{-iw_k t} \widetilde\psi_-(k,0) ~,
\end{equation}
where $\widetilde\psi_\pm(k,0)$ are the projections of the initial amplitude
distribution along $e_\pm$. The amplitude distribution in the position space
is given by the inverse Fourier transform of $\widetilde\psi(k,t)$. While we
are unable to evaluate it exactly, many properties of the quantum random
walk can be extracted numerically as well as by suitable approximations.

A walk starting at the origin satisfies $\psi_{\rm o}(n,0)=\delta_{n,0}$.
This walk is asymmetric because our definitions treat even and odd sites
differently. We can construct a symmetric walk, using the initial condition
$\psi_{\rm s}(n,0)=(\delta_{n,0}+i\delta_{n,1})/\sqrt{2}$. The resultant
probability distribution is then symmetric under $n\leftrightarrow(1-n)$.
(Real and imaginary components of the amplitude distribution evolve
independently because we have chosen the evolution operator to be real.)
For both these initial conditions, by construction, the quantum random
walk remains within the interval $[-2t+1,2t]$ after $t$ time steps.

The escape probability of the quantum random walk can be calculated by
introducing a fully absorbing wall, say between $n=0$ and $n=-1$.
Mathematically, this absorbing wall amounts to a projection operator for
$n\ge0$. The unabsorbed part of the walk is given by
\begin{eqnarray}
\psi(n,t+1) &=& P_{n\ge0}U_e U_o~\psi(n,t) ~,\\
            &=& U_e U_o~\psi(n,t) - {1\over2}\delta_{n,-1}(\psi(0,t)+\psi(1,t)) ~,
\nonumber
\end{eqnarray}
with the absorption probability,
\begin{equation}
P_{\rm abs}(t) = 1 - \sum_{n\ge0} |\psi(n,t)|^2 ~.
\end{equation}

All these variations in initial and boundary conditions are easy to implement
numerically, and examples are shown in Fig.2. We have used such simulations
to study various properties of the quantum random walk.

For large $t$, a good approximation to the probability distributions can be
obtained by the stationary phase method \cite{nayak,qwalk1}.
The smoothed probability distribution for the symmetric walk, obtained by
replacing the highly oscillatory terms by their mean values, is
\begin{equation}
\label{smooth}
|\psi_{\rm s}|_{\rm smooth}^2 = {4t^2 \over \pi\sqrt{4t^2-2n^2}~(4t^2-n^2)} ~.
\end{equation}
(Here, the $n\leftrightarrow(1-n)$ symmetry can be restored by replacing
$n$ by $(n-{1\over2})$.)
As shown in the top part of Fig.2, it represents the average behavior of the
distribution very well. Its low order moments are easily calculated to be,
\begin{eqnarray}
\label{moment0}
\int_{n=-\sqrt{2}t}^{\sqrt{2}t} |\psi_{\rm s}|_{\rm smooth}^2 dn &=& 1 ~, \\
\int_{n=-\sqrt{2}t}^{\sqrt{2}t} |n| \cdot |\psi_{\rm s}|_{\rm smooth}^2 dn &=& t ~, \\
\int_{n=-\sqrt{2}t}^{\sqrt{2}t} n^2 |\psi_{\rm s}|_{\rm smooth}^2 dn
     &=& 2(2-\sqrt{2})t^2 ~.
\label{moment2}
\end{eqnarray}

\begin{figure}[b]
\epsfxsize=9truecm
\centerline{\epsfbox{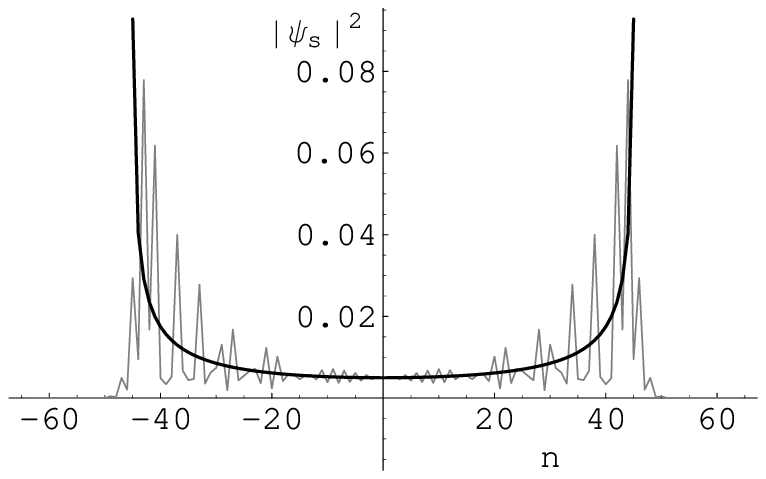}}
\vskip -1cm
\epsfxsize=9truecm
\centerline{\epsfbox{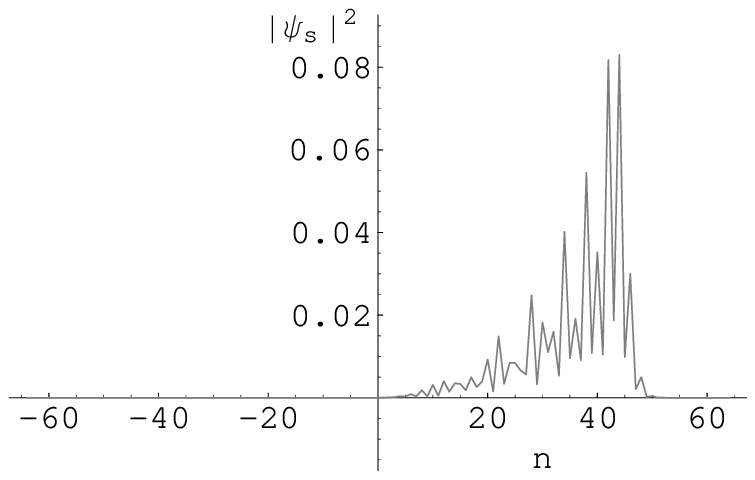}}
\vskip -1cm
\caption{Probability distribution after $32$ time steps for the symmetric
quantum random walk $\psi_{\rm s}$. In the top figure, the dark curve
denotes the smoothed distribution of Eq.(\ref{smooth}). The bottom figure
shows the distribution in the presence of an absorbing wall on the left
side of $n=0$.}
\end{figure}

\subsection{Results}

The following properties of the quantum random walk are easily deduced
\cite{qwalk1}:\\
$\bullet$ 
The probability distribution is double-peaked with maxima approximately at
$\pm\sqrt{2}t$. The distribution falls off steeply beyond the peaks, while
it is rather flat in the region between the peaks. With increasing $t$,
the peaks become more pronounced, because the height of the peaks decreases
more slowly than that for the flat region. The location of the peaks is in
accordance with the propagation speed, $|v|=1/\sqrt{2}$, once we take in to
account the fact that a single step of our walk is a product of two nearest
neighbor operators, $U_e$ and $U_o$.\\
$\bullet$
The size of the tail of the amplitude distribution is limited by
$(\epsilon t)^{-1} \sim t^{-1/3}$, which gives
$\Delta n_> = \Delta(\epsilon t) = O(t^{1/3})$.
On the inner side, the width of the peaks is governed by
$|\omega_k^{''} t|^{-1/2} \sim t^{-1/3}$. For $|n|=(\sqrt{2}-\delta)t$,
this gives $\Delta n_< = \Delta(\delta t) = O(t^{1/3})$.
The peaks therefore make a negligible contribution to the probability
distribution, $O(t^{-1/3})$.\\
$\bullet$
Rapid oscillations contribute to the probability distribution (and hence
to its moments) only at subleading order. They can be safely ignored in
an asymptotic analysis, retaining only the smooth part of the probability
distribution.\\
$\bullet$
The quantum random walk spreads linearly in time, with a speed smaller
by a factor of $\sqrt{2}$ compared to a directed walk. This speed is a
measure of its mixing behavior and hitting probability. The probability
distribution is qualitatively similar to a uniform distribution over the
interval $[-\sqrt{2}t,\sqrt{2}t]$. In particular, the $m^{\rm th}$ moment
of the probability distribution is proportional to $t^m$. This behaviour
is in sharp contrast to that of the classical random walk. The classical
random walk produces a binomial probability distribution, which in the
symmetric case has a single peak centered at the origin and variance
proportional to $t$. The linear spread in time of our quantum random walk
is achieved even when $\psi$ has 50\% probability to stay put at the same
location at every step, as can be seen from Eqs.(\ref{evolnocoin_o},%
\ref{evolnocoin_e}). This means that our walk is more directed and less
of a zigzag.\\
$\bullet$
Above properties agree with those obtained in Refs.\cite{nayak,watrous}
for a quantum random walk with a coin-toss instruction (extra factors
of $2$ appear in our results because of difference in our conventions),
demonstrating that the coin offers no advantage in this particular set up.
Essentially, we have absorbed the two states of the coin in to the even/odd
site label at no extra cost. By making the coin states part of the position
space, we have eliminated quantum entanglement between the coin and the
position degrees of freedom completely---only superposition representing
the amplitude distribution survives \cite{entangle}.
Such a reorganisation would be a tremendous advantage in any practical
implementation of the quantum random walk, because quantum entanglement is
highly fragile against environmental disturbances while mere superposition
is much more stable. The cost for gaining this advantage is the loss of
short distance homogeneity---translational invariance holds in steps of
$2$ instead of $1$.\\
$\bullet$
Comparison of the numerically evaluated probability distributions in Fig.2,
without and with the absorbing wall, shows that the absorbing wall disturbs
the evolution of the walk only marginally. The probability distribution in
the region close to $n=0$ is depleted as anticipated, while it is a bit of
a surprise that the peak height near $n=\sqrt{2}t$ increases slightly.
As a result, the escape speed from the wall is little higher than the
spreading speed without the wall. Overall, the part of the quantum random
walk going away from the absorbing wall just takes off at a constant speed,
hardly ever returning to the starting point. Again, this behavior is in a
sharp contrast to that of the classical random walk, which always returns
to the starting point, sooner or later. We also find that the first two time
steps dominate absorption, $P_{\rm s,abs}(t=1)=0.25$ and $P_{\rm s,abs}(t=2)
=0.375$, with very little absorption later on. Asymptotically, the net
absorption probability approaches $P_{\rm s,abs}(\infty) \approx 0.4098$
for the symmetric walk. This value is smaller than the corresponding result
$P_{\rm abs}(\infty)=2/\pi$ for the symmetric quantum random walk with a
coin-toss instruction \cite{watrous}.

\section{Quantum Random Walk\break on a Hypercubic Lattice}

\subsection{$2$-dim Lattice}

Next let us consider the situation for $d=2$. The partitioned free
Hamiltonian is given by
\begin{eqnarray}
H_o |x,y\rangle &=& -{i \over 2}\Big[ (-1)^x |x+(-1)^x,y\rangle \nonumber\\
                & &             + (-1)^{x+y} |x,y+(-1)^y\rangle \Big] ~, \\
H_e |x,y\rangle &=&  {i \over 2}\Big[ (-1)^x |x-(-1)^x,y\rangle \nonumber\\
                & &             + (-1)^{x+y} |x,y-(-1)^y\rangle \Big] ~, \\
H |x,y\rangle   &=& (H_o+H_e)|x,y\rangle \nonumber\\
                &=& -{i \over 2}\Big[ |x+1,y\rangle - |x-1,y\rangle \nonumber\\
                & &         + (-1)^x (|x,y+1\rangle - |x,y-1\rangle) \Big] ~.
\end{eqnarray}
More explicitly, the $4\times4$ blocks of the Hamiltonian are:
\begin{eqnarray}
H_o^B &=& -{i \over 2}\pmatrix{ 0&      -1&      -1&\hfill 0 \cr
                                1&\hfill 0&\hfill 0&\hfill 1 \cr
                                1&\hfill 0&\hfill 0&      -1 \cr
                                0&      -1&\hfill 1&\hfill 0 \cr}
                      ~\matrix{ 00 \cr 10 \cr 01 \cr 11 \cr} \\
      &=& -{1 \over 2}( I \otimes \sigma_2 + \sigma_2 \otimes \sigma_3) ~,
\label{2dHoB}
\end{eqnarray}
where the column on the right denotes the vertices of the elementary
square on which $H_o^B$ operates. Similarly, $H_e^B=-H_o^B$, when
operating on the square with vertices \{00,-10,0-1,-1-1\}. Noting that
$H_o^2 = H_e^2 = {1 \over 2}I$, the block-diagonal matrices are easily
exponentiated to
\begin{equation}
U_{o(e)} = cI - is\sqrt{2}H_{o(e)} ~,~~ |c|^2 + |s|^2 = 1 ~.
\end{equation}
The parameter $c$ (or $s$) is to be tuned to achieve the fastest diffusion
across the lattice.

\begin{figure}[b]
\vskip -2cm
\epsfxsize=8cm
\centerline{\epsfbox{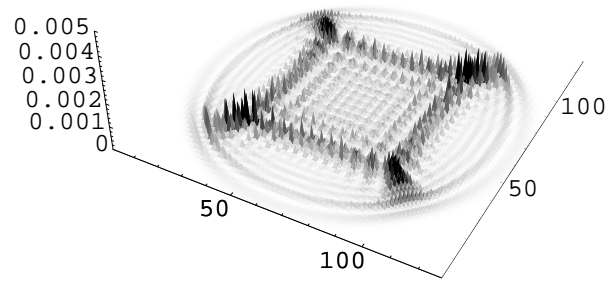}}
\caption{Probability distribution for the quantum random walk on a
two-dimensional $128\times128$ grid after $61$ time steps. A symmetric start,
i.e. $\psi(x,y,0) =(\delta_{x0}\delta_{y0}+i\delta_{x1}\delta_{y1})/\sqrt{2}$,
was used. Darker shades on the grey-scale indicate larger probability.}
\end{figure}

The quantum random walk with the Dirac operator spreads on a
two-dimensional grid as illustrated in Fig.3. The continuum Dirac
Hamiltonian has exact rotational symmetry, and that survives to an
extent even after discretisation on a hypercubic lattice. After a point
start, the random walk spreads essentially isotropically at distances
much larger than the lattice spacing, while the hypercubic symmetry
governs the random walk pattern at shorter distances. Of course, the
hypercubic symmetry would be exact for a $d$-dim random walk constructed
as a tensor product of $d$ one-dimensional random walks.

\subsection{$3$-dim Lattice}

Next let us look at the situation for $d=3$. The partitioned free
Hamiltonian is given by
\begin{eqnarray}
H_o|x,y,z\rangle&=& -{i\over2}\Big[ (-1)^x   |x+(-1)^x,y,z\rangle \nonumber\\
                & &           + (-1)^{x+y}   |x,y+(-1)^y,z\rangle \\
                & &           + (-1)^{x+y+z} |x,y,z+(-1)^z\rangle \Big] ~,
\nonumber\\
H_e|x,y,z\rangle&=&  {i\over2}\Big[ (-1)^x   |x-(-1)^x,y,z\rangle \nonumber\\
                & &           + (-1)^{x+y}   |x,y-(-1)^y,z\rangle \\
                & &           + (-1)^{x+y+z} |x,y,z-(-1)^z\rangle \Big] ~,
\nonumber\\
H |x,y,z\rangle &=& (H_o+H_e)|x,y,z\rangle \nonumber\\
                &=& -{i\over2}\Big[|x+1,y,z\rangle-|x-1,y,z\rangle\nonumber\\
                & &  + (-1)^x     (|x,y+1,z\rangle-|x,y-1,z\rangle\\
                & &  + (-1)^{x+y} (|x,y,z+1\rangle-|x,y,z-1\rangle)\Big] ~.
\nonumber
\end{eqnarray}
More explicitly, the $8\times8$ blocks of the Hamiltonian are:
\begin{eqnarray}
H_o^B &=& -{i \over 2}\pmatrix{
      0&      -1&      -1&\hfill 0&      -1&\hfill 0&\hfill 0&\hfill 0 \cr
      1&\hfill 0&\hfill 0&\hfill 1&\hfill 0&\hfill 1&\hfill 0&\hfill 0 \cr
      1&\hfill 0&\hfill 0&      -1&\hfill 0&\hfill 0&\hfill 1&\hfill 0 \cr
      0&      -1&\hfill 1&\hfill 0&\hfill 0&\hfill 0&\hfill 0&      -1 \cr
      1&\hfill 0&\hfill 0&\hfill 0&\hfill 0&      -1&      -1&\hfill 0 \cr
      0&      -1&\hfill 0&\hfill 0&\hfill 1&\hfill 0&\hfill 0&\hfill 1 \cr
      0&\hfill 0&      -1&\hfill 0&\hfill 1&\hfill 0&\hfill 0&      -1 \cr
      0&\hfill 0&\hfill 0&\hfill 1&\hfill 0&      -1&\hfill 1&\hfill 0 \cr}
~\matrix{ 000\cr 100\cr 010\cr 110\cr 001\cr 101\cr 011\cr 111\cr} \nonumber\\
      &=& -{1 \over 2}( I \otimes I \otimes \sigma_2
                      + I \otimes \sigma_2 \otimes \sigma_3
                      + \sigma_2 \otimes \sigma_3 \otimes \sigma_3) ~,
\label{3dHoB}
\end{eqnarray}
with the column on the right indicating the vertices of the elementary cube
on which $H_o^B$ operates. Likewise, $H_e^B=-H_o^B$, when operating on the
elementary cube with vertices \{000,-100,0-10,-1-10,00-1,-10-1,0-1-1,\break
-1-1-1\}. With $H_o^2 = H_e^2 = {3\over4}I$, the block-diagonal matrices
exponentiate to
\begin{equation}
U_{o(e)} = cI - is{2\over\sqrt{3}}H_{o(e)} ~,~~ |c|^2 + |s|^2 = 1 ~.
\end{equation}
Again $c$ (or $s$) is a parameter to be tuned to achieve the fastest
diffusion across the lattice.

\subsection{$d$-dim Lattice}

We can now observe a pattern in the explicit results for $d=1,2,3$ above.
The $2^d\times2^d$ blocks of the Hamiltonian can be written as sums of tensor
products of Pauli matrices. As suggested by Eqs.(\ref{2dHoB},\ref{3dHoB}),
\begin{equation}
H_o^B = -{1\over2} \sum_{j=1}^d
      I^{\otimes (d-j)} \otimes\sigma_2\otimes \sigma_3^{\otimes (j-1)} ~,
\end{equation}
and $H_e^B = -H_o^B$ when operating on the hypercube with coordinates
flipped in sign. The block-diagonal matrices satisfy
$H_o^2 = H_e^2 = {d \over 4}I$, and exponentiate to 
\begin{equation}
U_{o(e)} = cI - is{2\over\sqrt{d}}H_{o(e)} ~,~~ |c|^2 + |s|^2 = 1 ~.
\end{equation}

\section{Search on a Hypercubic Lattice\break using the Dirac Operator}

\subsection{Strategy}

A clear advantage of quantum random walks is their linear spread in time,
compared to square-root spread in time for classical random walks.
So they are expected to be useful in problems requiring fast hitting times.
Several examples of this nature have been explored in graph theoretical and
sampling problems (see Refs.\cite{kempe,ambainis} for reviews).
Here we consider the particular case of using the quantum random walk to
find a marked vertex on a hypercubic lattice (see also
Refs.\cite{gridsrch1,gridsrch2}).

Consider a $d$-dim hypercubic lattice with $N=L^d$ vertices, one of which
is marked. The quantum algorithmic strategy for the search process is to
construct a Hamiltonian evolution, where the kinetic part of the Hamiltonian
diffuses the amplitude distribution all over the lattice while the potential
part of the Hamiltonian attracts the amplitude distribution towards the
marked vertex \cite{grover_strategy}.
The optimisation criterion is to concentrate the amplitude distribution
towards the marked vertex as quickly as possible. In his algorithm, Grover
constructed a global operator that allows diffusion from any vertex to any
other vertex in just one step. Under different circumstances, when diffusion
is restricted to be ultra-local (i.e. one can only go from a vertex to its
neighbours in one step), one must find an appropriate diffusion operator
that provides fast propagation of spatial modes. Obviously, the Dirac
operator is better suited to this task than the Laplacian operator.

To search for a marked vertex, say the origin, we need to attract the
quantum random walk towards it. This can be accomplished by adding a
potential to the free Hamiltonian,
\begin{equation}
V = V_0 ~\delta_{\vec{x},0} ~.
\end{equation}
Exponentiation of this potential produces a phase change for the amplitude
at the marked vertex. It is optimal to choose the magnitude of the potential
to make the phase maximally different from $1$, i.e. $e^{-iV_0\tau}=-1$,
whereby the phase becomes a reflection operator (binary oracle),
\begin{equation}
R = I - 2 |\vec{0}\rangle\langle\vec{0}| ~.
\end{equation}
The search algorithm alternates between the diffusion and the reflection
operators, yielding the evolution
\begin{equation}
\psi(\vec{x};t_1,t_2) = [W^{t_1} R]^{t_2} \psi(\vec{x};0,0) ~.
\end{equation}
Here $t_2$ is the number of oracle calls, and $t_1$ is the number of random
walk steps between the oracle calls. Both have to be optimised, in addition
to $c$ and depending on the size and dimensionality of the lattice, to find
the quickest solution to the search problem.

Fastest search amounts to finding the shortest unitary evolution path
between the initial state, typically chosen as the uniform superposition
state $|s\rangle=\sum_x|\vec{x}\rangle/\sqrt{N}$, and the marked state
$|\vec{0}\rangle$. This path is a circular arc (geodesic) from $|s\rangle$
to $|\vec{0}\rangle$. With the random walk diffusion operator $W$, evolution
of the state $|\psi\rangle$ does not remain restricted to the two-dimensional
subspace formed $|s\rangle$ and $|\vec{0}\rangle$. Thus to optimise our
algorithm, we need to tune the parameters so as to\\
(a) maximise the projection of the state $|\psi\rangle$ on to the
two-dimensional $|s\rangle-|\vec{0}\rangle$ subspace, and\\
(b) maximise the angle of rotation by the operator $W^{t_1}R$, for the
projected component of $|\psi\rangle$ in the $|s\rangle$-$|\vec{0}\rangle$
subspace.\\
We have explored this optimisation numerically.

\begin{figure}[b]
\vskip -1cm
\epsfxsize=8cm
\centerline{\epsfbox{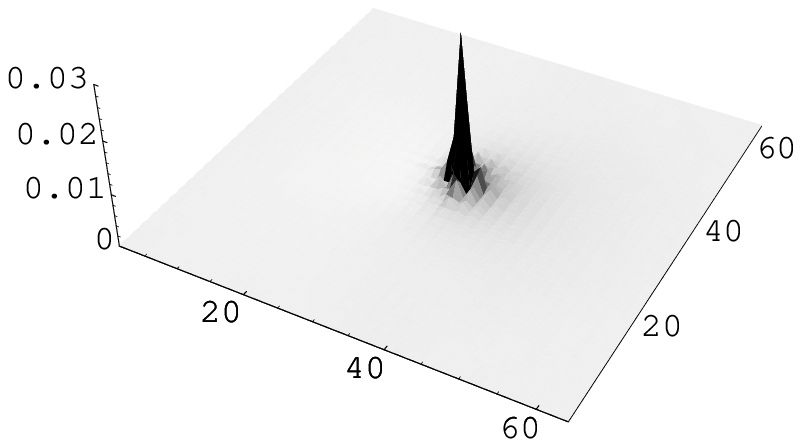}}
\vskip -1cm
\epsfxsize=8cm
\centerline{\epsfbox{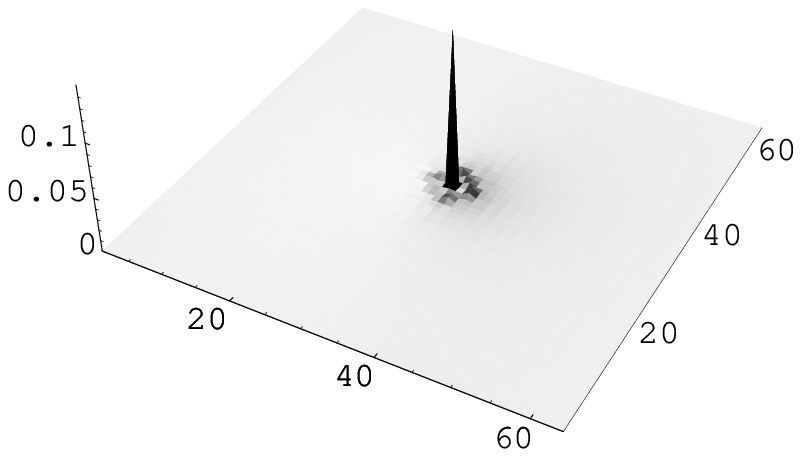}}
\caption{Probability distribution for the quantum random walk search problem
on a two-dimensional $64\times64$ grid, at the instance when the probability
at the marked vertex attains its largest value. The number of random walk
steps are $t_1=1$ (top) and $t_1=3$ (bottom).}
\end{figure}

\subsection{Numerical Results}

\begin{figure*}[t]
\centerline{\epsfxsize=9.5cm \epsfbox{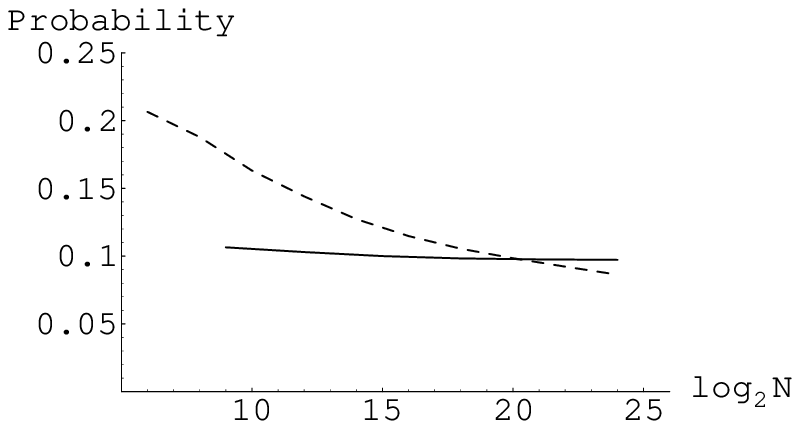}
            \epsfxsize=8.5cm \epsfbox{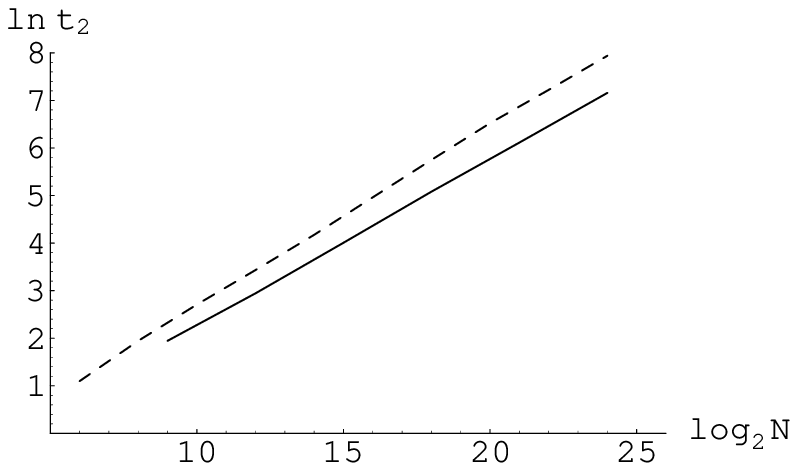}}
\caption{The peak probability at the marked vertex for the quantum random
walk search problem (left), and the number of time steps required to reach
it (right), as a function of the database size. The dashed and continuous
curves correspond to $d=2$ and $d=3$ respectively. $c=1/\sqrt{2}$ and
$t_1=3$ were used, with $N$ ranging from $2^6$ to $2^{24}$.}
\end{figure*}

We carried out computer simulations of the quantum random walk search problem
with a single marked vertex, for $d=2,3$. The algorithm was optimised by
tuning the parameters $c$ and $t_1$, so as to minimise the number of oracle
calls $t_2$ required to find the marked vertex. The following is a summary
of our observations:\\
$\bullet$
An unbiased search starts with a uniform probability distribution
over the whole lattice. Thereafter, the probability at the marked vertex
goes through periodic cycles of rise and fall as a function of time step.
It is crucial to stop the algorithm at the right instance to find the
marked vertex with a significant probability.\\
$\bullet$
Our best results are obtained with $c=1/\sqrt{2}$ and $t_1=3$.
In this case, the probability at the marked vertex reaches its largest
value, and $t_2$ achieves its smallest value. With these parameters, the
probability at the marked vertex shows a periodic sinusoidal behaviour,
which persists for more than 30 cycles without any visible deviation.
Also, apart from the uniform background, the probability distribution
shows a sharp single-point delta function at the marked vertex. These
features indicate that the walk evolves largely in the two-dimensional
subspace formed by the uniform state and the marked state.\\
$\bullet$
For $c<1/\sqrt{2}$, the walk diffuses more slowly, and $t_2$ increases.
For $c>1/\sqrt{2}$, the probability at the marked vertex loses its periodic
sinusoidal behaviour, suggesting that the walk no longer remains confined
to the two-dimensional subspace. For the optimal choice $c=1/\sqrt{2}$, the
probability of the walk remaining at the same vertex equals that for moving
to a neighbouring vertex, which corresponds to the most efficient mixing
between odd and even sublattices.\\
$\bullet$
For $t_1<3$, the probability distribution spreads out instead of being a
delta function at the marked vertex, as illustrated in Fig.3. This decreases
the peak probability at the marked vertex. Moreover, $t_2$ increases. For
$t_1>3$, the probability at the marked vertex loses its sinusoidal behaviour,
again with a decrease in the peak probability. In both cases, the changes
indicate that the walk is drifting out of the two-dimensional subspace.
An appropriate choice of $t_1$ is thus crucial to keep the walk close to
the two-dimensional subspace.\\
$\bullet$
For the $2$-dim walk, the largest probability at the marked vertex
is predicted to be $O(1/\log N)$, which occurs after $O(\sqrt{N \log N})$
time steps \cite{gridsrch1,gridsrch2}.
To make the marked vertex probability $O(1)$, an amplitude amplification
procedure is required \cite{brassard},
and the overall search algorithm scales as $O(\sqrt{N} \log N)$.
Our numerical results, shown in Fig.5, are consistent with these
expectations. Simple fits provide the parametrisations:
\begin{eqnarray}
O(1/\log N) &\longrightarrow& 2.12/ \log_2 N ~,\\
O(\sqrt{N \log N}) &\longrightarrow& 0.137 \sqrt{N \log_2 N} ~. \nonumber
\end{eqnarray}
$\bullet$
For the walk in more than two dimensions, the largest probability at the
marked vertex is predicted to be $O(1)$, which occurs after $O(\sqrt{N})$
time steps \cite{gridsrch1,gridsrch2}.
Our numerical results for the $3$-dim walk, also displayed in Fig.5,
agree with these scaling rules. Simple fits provide the parametrisations:
\begin{equation}
O(1) \longrightarrow 0.0969 ~,~~ O(\sqrt{N}) \longrightarrow 0.313 \sqrt{N} ~.
\end{equation}

These results demonstrate that our quantum random walk algorithm achieves
the optimal scaling behaviour for the problem of finding a marked vertex
on a hypercubic lattice. Thus our quantum random walk, based on the Dirac
operator and not containing a coin toss instruction, is no less effective
in its diffusion properties than the earlier quantum random walks that use
a coin toss instruction.


\begin{thebibliography}{99}
\bibitem{massterm}{In general, energy includes a mass term. The mass
         operator behaves like the $\vec{k}=0$ mode, and changes the
         phase of the quantum amplitude. But it does not contribute
         to diffusion, and so we leave the mass term out.}
\bibitem{spinorsize}{This is in spite of the fact that Dirac spinors
         require more than two components for $d>2$. For example, in
         $3$-dimensions, a Rubik's cube can have the central cube
         labeled ``o'' and the eight corner cubes labeled ``e''.}
\bibitem{phaseshift}{This transformation is a phase-shift of $\pi/2$ in
         momentum space. It changes $\cos(k)$ terms associated with second
         derivatives to $\sin(k)$ terms associated with first derivatives.}
\bibitem{entangle}{Entanglement is always defined with respect to a specific
division of the whole system in to its parts. If the division scheme is
altered, entanglement can change.}
\end{thebibliography}
\end{document}